\documentstyle[12pt]{article}
\def\bqn{\begin{equation}}
\def\nqn{\end{equation}}

\begin{document}

\begin{titlepage}
\vspace{2.5cm}
\begin{center}{\large \vspace{4.0cm} \bf Dynamical Symmetries and
Nambu Mechanics }\\
\vspace{1cm}
Rupak Chatterjee \footnotemark  \\
\footnotetext{e-mail: rupak@max.physics.sunysb.edu}
\vspace{1cm}
Department of Physics, State University of New York at Stony Brook \\
Stony Brook, NY 11794-3800, USA \\
\vspace{0.5cm}
January, 1995
\vspace{3cm}

\begin{abstract}

It is shown that several Hamiltonian systems
possessing dynamical or hidden symmetries can be
realized within the framework of Nambu's
generalized mechanics.
Among such systems are the SU(n)-isotropic
harmonic oscillator and the SO(4)-Kepler problem.
As required by the
formulation of Nambu dynamics, the integrals of
motion for these
systems necessarily become the so-called
generalized Hamiltonians.
Furthermore, in most of these problems, the definition
of these generalized Hamiltonians is not unique.

\end{abstract}

\end{center}
\end{titlepage}

\section{Introduction}

More than two decades ago, Nambu [1] proposed a generalization
to classical Hamiltonian mechanics.
In his formalism, he replaced the usual pair of canonical variables found
in Hamiltonian mechanics with a triplet of coordinates in
an odd dimensional phase space.
Furthermore, he formulated his dynamics via a
ternary operation, the Nambu bracket, as opposed to the
usual binary Poisson bracket.
Yet, the fundamental principles of a canonical form
of Nambu's generalized mechanics, similar to the invariant geometrical
form of Hamiltonian mechanics, has only recently been given [2].
The re-emergence of this little known theory is possibly due to its
relevence to the recent mathematical structures having their
basis in Hamiltonian mechanics such as the Poisson-Lie group,
quantum groups, and the Yang-Baxter equation. Since the basic idea
of Nambu mechanics is to extend the usual binary operation on phase space
to multiple operations of higher order, this theory may also give
some insights into the theory of higher order algebraic structures
and their possible physical significance.

The fact that the development of Nambu mechanics is still at the
preliminary stages can be seen by the relatively few known examples
of dynamical systems which admit a Nambu-type formulation.
Nambu himself came up with only one example; the Euler equations
for the angular momentum of a rigid body in three dimensions [1].
Another (somewhat exotic) example is that of Nahm's system of equations
in the theory of static SU(2)-monopoles [3], [4]. A third example
was found in [2].
Here, we
present four new examples of systems taking on the Nambu form. These
systems share the common property of possessing dynamical or hidden
symmetries resulting in extra integrals of motion beyond those
needed for complete integrability. It is felt that these new examples
may help in further understanding the elements of Nambu's theory
such as its algebraic structure and its (possible)
quantization. We begin by stating the basic facts of Nambu
mechanics leaving all details to the recent comprehensive study
by Takhtajan [2].

\vspace{1.0cm}

\section {Nambu Mechanics}

For comparisons sake, let us first review certain fundamental definitions
and results of Hamiltonian mechanics (see [5]). Let $M$ denote
a smooth manifold of finite dimension and $C^{\infty }(M)$ the
algebra of smooth real-valued functions on $M$. A {\em Poisson
Bracket} on $M$ is a ${\rm I\kern-.20em R}$-bilinear map
\bqn
\{ ~, ~ \} ~:~ C^{\infty }(M) \times  C^{\infty }(M) \rightarrow
C^{\infty }(M),
\nqn
such that $\forall ~f_1, f_2,$ and $f_3 \in C^{\infty }(M)$,
\bqn
\{ f_1 , f_2 \} = -\{ f_2 , f_1 \} ~~~~(skew-symmetry),
\nqn
\bqn
\{ f_1 f_2 , f_3 \} = f_1 \{ f_2 , f_3 \} + \{ f_1 , f_3 \} f_2
    ~~~~(Leibniz ~rule),
\nqn
\bqn
\{ f_1 , \{ f_2 , f_3 \} \} + \{ f_2 , \{ f_3 , f_1 \} \}
+ \{ f_3 , \{ f_1 , f_2 \} \} = 0 ~~~~(Jacobi ~identity).
\nqn
By property (3), it is clear that $\forall ~h \in C^{\infty }(M)$,
the map $ f \rightarrow \{ f , h \} $ is a derivation of $C^{\infty }(M)$.
Thus, one may define a vector field $X_h $ such that $X_h (f) = \{ f , h \}$,
$\forall ~f \in C^{\infty }(M)$. In Hamilton's formulation of
dynamics, one defines a very special vector field $X_H $ on the phase
space $M$ via a Hamiltonian function $H \in C^{\infty }(M)$ such that
\bqn
X_H (f) = \{ f , H \} = \frac{df}{dt} , ~~\forall ~f \in C^{\infty }(M).
\nqn
This, of course, produces Hamiltons' equations of motion.

The Poisson bracket given above is explicitly defined
by a $C^{\infty }$ tensor field
$\omega \in \wedge ^2 TM$, an element of the square of the tangent
bundle $TM$ of the phase space $M$, such that
\bqn
\{ f , g \} = \omega (df,dg) =  \omega ^{ij} \frac {\partial{f}}
{\partial{x^i}} \frac {\partial{g}}{\partial{x^j}},
\nqn
where $(x^i)$ are some local coordinates on $M$ and we use the Einstein
summation convention. For the purposes of regular
Hamiltonian dynamics, $\omega ^{ij}$ is a constant antisymmetric
two-tensor. Outside of Hamiltonian mechanics, there exists
linear Poisson structures such that if $(x_1 , x_2 , ... , x_n )$
is a basis for a Lie algebra, the Poisson bracket is given by
\bqn
\{ f , g \} = [x_i , x_j ]_{Lie} ~ \frac {\partial{f}}{\partial{x^i}}
\frac {\partial{g}}{\partial{x^j}},
\nqn
$[~,~]_{Lie} $ denoting the Lie algebra bracket. A more modern example
is that of a {\em Poisson-Lie group} where the phase space $M$ is some
Lie group $G$ such that the group multiplication of $G$ is compatible,
in some sense, with the Poisson-structure on $G$ (for a precise definition,
see [6]). A complete description of such structures (Poisson-Lie Manifolds)
may be found in the monograph [7].

Finally, the dynamical picture resulting from a solution to (5)
produces a phase space (time) flow
\bqn
x \longmapsto T_t (x), ~~x \in M
\nqn
and an evolution operator
\begin{eqnarray}
U_t : C^{\infty }(M) & \rightarrow & C^{\infty }(M), \nonumber \\
U_t (f)(x) & = & f(T_t (x)),
\end{eqnarray}
where $x \in M,~f \in C^{\infty }(M)$.
For consistency, $U_t$ must preserve the two algebraic operations defined
on $C^{\infty }(M)$, the usual multiplication of functions and the
Poisson bracket. This requirement of $U_t$ to be an algebra isomorphism
($U_t (f_1 f_2 ) = U_t (f_1) U_t (f_2)$ and $U_t (\{ f_1 , f_2 \} ) =
\{ U_t (f_1 ) , U_t (f_2 ) \} $) is equivalent to the properties
(3) and (4) of the Poisson bracket. A similar requirement will lead
to a certain fundamental identity in Nambu mechanics.

Nambu's generalization of mechanics is based upon a higher order
($n~>~2$) algebraic structure defined on a (possibly odd
dimensional) phase space $M$. A {\em Nambu
Bracket} of order $n$ on a manifold $M$ is an ${\rm I\kern-.20em R}$
-multilinear map
\bqn
\{ ~,~\ldots ~,~ \}~:~[C^{\infty }(M)]^{\otimes n} \rightarrow C^{\infty }(M)
\nqn
such that $\forall f_1 , f_2 , \ldots , f_{2n-1} \in C^{\infty }(M)$,
\bqn
\{ f_1, \ldots ,f_n \}=(-1)^{P(\sigma)}\{ f_{\sigma(1)}, \ldots ,
f_{\sigma(n)} \},
\nqn
\bqn
\{ f_1 f_2, f_3, \ldots ,f_{n+1} \}=
f_1 \{f_2, f_3, \ldots , f_{n+1} \} +
\{ f_1, f_3, \ldots, f_{n+1} \} f_2,
\nqn
and
\begin{eqnarray}
\{ \{ f_1, \ldots , f_{n-1}, f_n \}, f_{n+1}, \ldots, f_{2n-1} \} +
\{ f_n, \{ f_1, \ldots, f_{n-1}, f_{n+1} \}, f_{n+2}, \ldots , f_{2n-1} \} \\
 +  \ldots + \{ f_n, \ldots ,f_{2n-2}, \{ f_1, \ldots , f_{n-1}, f_{2n-1} \}\}
 =  \{ f_1, \ldots , f_{n-1}, \{ f_n, \ldots , f_{2n-1} \}\}, \nonumber
\end{eqnarray}
where $\sigma \in S_n$ (the permutation group) and $P$ is the parity
of the permutation $\sigma $.
Conditions (11) and (12) are the familiar skew-symmetric and derivation
properties found for the Poisson Bracket. On the other hand,
(13) is a generalized Jacobi identity
called the Fundamental Identity (FI). The FI was first introduced by
Takhtajan [2] and by M. Flato and C. Fronsdal. It was also independently
found by Sahoo and Valsakumar [8], [9].

Now, analagous to (5), Nambu dynamics is determined by a special
Nambu-Hamiltonian vector field $X_{NH}$ given by
\bqn
X_{NH}(f) = \{ f ,~ H_1 ,~ \ldots ~,~H_{n-1} \} = \frac {df}{dt},
\nqn
$\forall ~f \in C^{\infty }(M)$ where $H_1 ,~ \ldots ,~H_{n-1}$ are the
generalized Hamiltonians of the system. A solution to the above equations
of motion produces an evolution operator $U_t$ as described
earlier by (8) and (9). As is the case in Hamiltonian mechanics, Nambu
dynamics is consistent if and only if $U_t$ is an isomorphism of the
above defined algebraic structure on $C^{\infty }(M)$. It can be shown
that the corresponding isomorphism requirement of
\bqn
U_t ( \{ f_1 ~,~\ldots ~,f_n  \}) = \{ U_t (f_1 ),~\ldots ,~
U_t ( f_n ) \}
\nqn
is equivalent to the FI (13) for the Nambu bracket. Furthermore,
the FI provides  a very important dynamical result. First of all,
a function $I \in C^{\infty }(M)$ is an {\em integral of motion}
if $\{ I,~H_1 ,~ \ldots ,~H_{n-1} \}=0$. Then, using the FI, one can
prove that the Nambu bracket of $n$ integrals of motion is also
an integral of motion (analogous to {\em Poisson's Theorem} in
Hamiltonian mechanics). Consequently, one can define the concept of
integrability in Nambu mechanics.

The Nambu bracket is explicitly generated by a {\em Nambu tensor field}
$\eta \in \wedge ^{n} TM$ such that
\bqn
\{ f_1, \ldots , f_n \} = \eta(df_1, \ldots, df_n) =
\eta_{i_1...i_n}(x) \frac{\partial}{\partial x_{i_1}} \wedge \ldots
\wedge \frac{\partial}{\partial x_{i_n}},
\nqn
where $(x_1 , x_2 , \ldots )$ are some local coordinates on $M$
and repeated indices are summed. The FI imposes
serious constraints on the Nambu tensor field $\eta $. Yet, for our
purposes, we only need to define the simplest type of Nambu bracket.
Let $M={\rm I\kern-.20em R}^{n}$ be our phase space with coordinates
$x_1,\ldots ,x_n$. Then the so-called `canonical' Nambu bracket is
\bqn
\{ f_1, \ldots , f_n \} = C~\frac {\partial (f_1, \ldots , f_n )}
{\partial (x_1,\ldots ,x_n )}
\nqn
where the right hand side is the Jacobian of the mapping
$(f_1, \ldots , f_n )~:~{\rm I\kern-.20em R}^{n} \rightarrow
{\rm I\kern-.20em R}^{n}$ and $C$ is a constant factor. This
bracket will be used throughout the rest of the paper.

\section {The SU(n)-Isotropic Harmonic Oscillator}

It is well known that the Hamiltonian for the n-dimensional simple
harmonic oscillator where all the frequencies have been set to one,
\bqn
H= \sum_{i=1}^{n}(p_{i}^2 + q_{i}^2)
\nqn
is invariant under the symmetry group SU(n) and has the
following integrals of motion,
\bqn
L_{ij} = q_i p_j - q_j p_i ,
\nqn
and
\bqn
A_{ij} = p_i p_j + q_i q_j ,
\nqn
where $i,j=1, \ldots ,n$.
Obviously, the $L_{ij}$'s are the angular momenta of the system
whereas the diagonal components of $A_{ij}$ are the individual
energies associated with the separate one-dimensional oscillations.
It is the off-diagonal components of $A_{ij}$ which provide the
hidden integrals of motion thus forcing all the trajectories
in phase space to lie on curves. That is, the $L_{ij}$'s
and the $A_{ij}$'s provide $(2n-1)$ independent integrals
of motion within the $2n$ dimensional phase space, and therfore all orbits
lie on one-dimensional manifolds. By Bertrand's theorem (see [10]),
these orbits are closed implying that the extra dynamical
integrals of motion are simple functions of the phase space coordinates.

Beginning with the two-dimensional case, if one defines the following
functions
\begin{eqnarray}
S_1 &=&  (A_{12}+A_{21})/2, \nonumber \\
S_2 &=&  (A_{22}-A_{11})/2, \\
S_3 &=& \frac{L_{12}}{2}, \nonumber
\end{eqnarray}
it is easy to verify that
\bqn
\{ S_i , S_j \} = \epsilon _{ijk} S_k
\nqn
which are simply the commutation relations for the Lie algebra
su(2). Furthermore, the Casimir function is related to the Hamiltonian
since
\bqn
S_{1}^2 + S_{2}^2 + S_{3}^2 = S^2 = \frac {H^2}{4}.
\nqn
The integrals of motion for this system are,
\begin{eqnarray}
I_1 &=& p_{1}^2 + q_{1}^2 = C_1 , \nonumber \\
I_2 &=& p_{2}^2 + q_{2}^2 = C_2 , \\
I_3 &=& q_1 p_2 - q_2 p_1 = C_3 , \nonumber \\
I_4 &=& p_1 p_2 + q_1 q_2 = C_4 , \nonumber
\end{eqnarray}
where the $C_i$'s are the constant values taken by the $I_i$'s.
Using these $I_i$'s as the generalized Hamiltonians, we can describe the
corresponding Nambu dynamics as follows. Consider the following Nambu
bracket,
\bqn
\{ f , I_1 , I_2 , I_3 \} = \left( \frac {-1}{2C_4} \right)
\frac {\partial (f , I_1 , I_2 , I_3 )}
{\partial (p_1 , p_2 , q_1 , q_2)},
\nqn
where, as before, the right hand side symbolizes the Jacobian
operation and $f$ is some function of the phase space coordinates.
Straightforward calculations will show that
this bracket in fact  produces all the correct equations
of motion, i.e. ${dp_1 }/{dt} = \{ p_1 , I_1 , I_2 , I_3 \} = -2q_1$, etc.
Note that since only three generalized Hamiltonians were needed,
$I_4$ was not used at all. Another definition, using $I_4$, and
giving the correct equations of motion is
\bqn
\{ f , I_1 , I_2 , I_4 \} = \left( \frac {1}{2C_3} \right)
\frac {\partial (f , I_1 , I_2 , I_4 )}
{\partial (p_1 , p_2 , q_1 , q_2)}.
\nqn
Even products of $I_i$'s can be used as generalized Hamiltonians.
For instance, the bracket
\bqn
\{ f , I_1 , I_2 , I_3 I_4 \} =
\frac {1}{2(C_{3}^{~2} - C_{4}^{~2})}
\frac {\partial (f , I_1 , I_2 , I_3 I_4 )}
{\partial (p_1 , p_2 , q_1 , q_2)}
\nqn
also works. By the derivation property (12) for Nambu brackets, it can be
checked that the alternate definitions (25)-(27) are consistant with
each other. For example, let $f=p_1 $ in (27) and add to the brackets
in (25), (26), and (27) the extra subscripts $I_3$, $I_4$, and $I_{34}$
respectively to distinguish them. Then

$$ \{  p_1 , I_1 , I_2 , I_3 I_4 \} _{I_{34}} = I_3 \{ p_1 , I_1 ,
I_2 , I_4 \} _{I_{34}} + \{ p_1 , I_1 , I_2 , I_3 \} _{I_{34}} I_4 $$

$$ = \frac{I_3 }{ 2(C_{3}^{~2} - C_{4}^{~2}) }
\frac{ \partial (p_1 , I_1 , I_2 , I_4 ) }
{ \partial (p_1 , p_2 , q_1 , q_2) } +
\frac{I_4 }{ 2(C_{3}^{~2}- C_{4}^{~2}) }
\frac{ \partial (p_1 , I_1 , I_2 , I_3 ) }
{ \partial (p_1 , p_2 , q_1 , q_2) } $$

$$ = \frac{2C_{3}^{~2} }{ 2(C_{3}^{~2} - C_{4}^{~2}) }
\{ p_1 , I_1 , I_2 , I_4 \} _{I_4 }
- \frac{2C_{4}^{~2} }{ 2(C_{3}^{~2} - C_{4}^{~2}) }
\{ p_1 , I_1 , I_2 , I_3 \} _{I_3 } $$

$$ = \frac{2C_{3}^{~2} }{ 2(C_{3}^{~2} - C_{4}^{~2}) } (-2q_1)
- \frac{2C_{4}^{~2} }{2(C_{3}^{~2} - C_{4}^{~2})} (-2q_1)
= ~-2q_1 $$
as needed. It is a staightforward exercise to verify that the above
calculation also holds for the remaining phase space variables.

By the above arguments and the fact that
$$ \frac{ \partial (I_1 , I_2 , I_3 , I_4 ) }
{ \partial (p_1 , p_2 , q_1 , q_2) }  = 0, $$
it is clear that one may choose any three of the four integrals
of motion (24) as the generalized Hamiltonians. One must simply
find the correct factors in front of the common Jacobian
term. Once these basic brackets have been established,
linear combinations such as $\{ f , I_1 , I_2 + I_3 , I_4 \} $ and
those of the type similar to that of (27) can easily be found
using the linearity and derivation properties of the Nambu
bracket. This extra flexibility not found in Hamiltonian mechanics
is obviously due to the multiple Hamiltonian structure of Nambu mechanics.
Finally, it is relatively straightforward to extend the above
arguments to higher dimensional systems. One simply uses the
integrals of motion given by (19) and (20) as the generalized
Hamiltonians and finds the correct constants related to these
integrals to multiply the Jacobian term in the definition
of the Nambu bracket.
Thus, the SU(n)-isotropic harmonic oscillator
is realizable as an $n^{th}$ order Nambu mechanical system.

\section {The SO(4)-Kepler Problem}

The well known Kepler Hamiltonian is
\bqn
H= \frac{\vec{p}^{~2}}{2} - \frac{1}{r},
\nqn
where $r=\sqrt{(x^2 + y^2 + z^2 )}$. Because $H$ possesses rotational
symmetry, the orbital angular momentum $\vec{L} = \vec{r} \times \vec{p}$
is an integral of motion. This rotational symmetry implies that the orbit
lies in some two dimensional plane, though it is not enough to ensure that
the orbit is closed. An extra dynamical symmetry must exist for a closed
orbit since the integrals of motion $H$ and $\vec{L}$ only reduce
the phase space to a two dimensional (as opposed to a one dimensional)
manifold. Such an integral was first discovered by Laplace
(but is called the Runge-Lenz vector in classical mechanics
or the Lenz-Pauli vector in quantum mechanics) and is given by
\bqn
\vec{A}=\vec{p} \times \vec{L} - \frac{\vec{r}}{r}.
\nqn
One can easily check that
$$ \{ A_i , L_j \} = \epsilon _{ijk} A_k $$
and
$$ \{ A_i , A_j \} = -\epsilon _{ijk} \left( \vec{p}^{~2} - \frac{2}{r}
\right) L_k = -2HL_k = -2EL_k , $$
where $E$ is the constant value taken by $H$. For bound state
problems ($E < 0$), one can define a new conserved vector $\vec{D}$
as
$$ \vec{D} = \frac{\vec{A}}{\sqrt{-2E}}. $$
Then one finds that the commutation relations reduce to
\begin{eqnarray*}
 \{ L_i , L_j \} &=& \epsilon _{ijk} L_k, \\
 \{ D_i , L_j \} &=& \epsilon _{ijk} D_k, \\
 \{ D_i , D_j \} &=& \epsilon _{ijk} L_k,
\end{eqnarray*}
which is the Lie algebra so(4).
(Note that for scattering problems where $E > 0 $,
one instead finds the Lie algebra so(3,1)).

Explicitly, we have
\bqn
H=\frac{p_1 ^2 + p_2 ^2 + p_3 ^2}{2} - \frac{1}{(q_1 ^2 +
q_2 ^2 + q_3 ^2)^{1/2}},
\nqn
and,
\begin{eqnarray}
I_1 &=& p_2 ( q_1 p_2 - q_2 p_1 ) - p_3 ( q_3 p_1 - q_1 p_3 ) -
\frac{q_1 }{(q_1 ^2 + q_2 ^2 + q_3 ^2)^{1/2}} = C_1,
\nonumber \\
I_2 &=& p_3 ( q_2 p_3 - q_3 p_2 ) - p_1 ( q_1 p_2 - q_2 p_1 ) -
\frac{q_2 }{(q_1 ^2 + q_2 ^2 + q_3 ^2)^{1/2}} = C_2,
\nonumber \\
I_3 &=& p_1 ( q_3 p_1 - q_1 p_3 ) - p_2 ( q_2 p_3 - q_3 p_2 ) -
\frac{q_3 }{(q_1 ^2 + q_2 ^2 + q_3 ^2)^{1/2}} = C_3,
\nonumber \\
I_4 &=& ( q_2 p_3 - q_3 p_2 ) = C_4, \\
I_5 &=& ( q_3 p_1 - q_1 p_3 ) = C_5, \nonumber \\
I_6 &=& ( q_1 p_2 - q_2 p_1 ) = C_6, \nonumber
\end{eqnarray}
with the relations
\begin{eqnarray}
I_1 ^{~2} + I_2 ^{~2} + I_3 ^{~2} &=& 1 + 2 H ( I_4 ^{~2}
+ I_5 ^{~2} + I_6 ^{~2} ), \nonumber \\
I_1 I_4 + I_2 I_5 + I_3 I_6 &=& 0,
\end{eqnarray}
where, as before, the $ C_i $'s are the constant values taken
by the $ I_i $'s.
Therefore, there exists only five independent constants of motion
as expected. As was the case for the harmonic oscillator, one
may choose any five of the above six
$ I_i $'s (or products thereof)
as the generalized Hamiltonians.
For example, one may define the
Nambu bracket for this system as
$$
\{ f , I_2 , I_3 , I_4 , I_5 , I_6 \} = \left( \frac {1}{C_4
(C_4 ^2 + C_5 ^2 + C_6 ^2 )} \right)
\frac {\partial (f , I_2 , I_3 , I_4 , I_5 , I_6 )}
{\partial (p_1 , p_2 , p_3 , q_1 , q_2 , q_3 )},
$$
or,
\bqn
\{ f , I_1 , I_3 , I_4 , I_5 , I_6 \} = \left( \frac {-1}{C_5
(C_4 ^2 + C_5 ^2 + C_6 ^2 )} \right)
\frac {\partial (f , I_1 , I_3 , I_4 , I_5 , I_6 )}
{\partial (p_1 , p_2 , p_3 , q_1 , q_2 , q_3 )},
\nqn
or,
$$
\{ f , I_1 , I_2 , I_3 , I_4 , I_5 \} = \left( \frac {-1}{C_3
(C_4 ^2 + C_5 ^2 + C_6 ^2 )} \right)
\frac {\partial (f , I_1 , I_2 , I_3 , I_4 , I_5 )}
{\partial (p_1 , p_2 , p_3 , q_1 , q_2 , q_3 )},
$$
etc. An interesting fact to note here is that we were unable
to (directly) incorporate the original Hamiltonian $H$ into
a form of the Nambu bracket.

\section{Two More Examples}

First of all, let us analyse a Hamiltonian related to the motion
of two vortices in an ideal incompressible fluid. A physical
description of this system may be found in the monograph [10].
The Hamiltonian and the corresponding integrals of motion are
\bqn
H=\ln {[(q_1 - q_2 )^2 + (p_1 - p_2 )^2 ]} = E,
\nqn
and,
\begin{eqnarray}
I_1 &=& q_1 + q_2 = C_1 , \nonumber \\
I_2 &=& p_1 + p_2 = C_2 ,\\
I_3 &=& p_1 ^2 + p_2 ^2 + q_1 ^2  + q_2 ^2 = C_3 . \nonumber
\end{eqnarray}
Since we have three independent integrals of motion
and a four dimensional phase space, (35) incorporates a dynamical
symmetry reducing the flow in phase space to a (not necessarily closed)
curve. This system, like the previous examples, has several
different Nambu brackets all producing the correct equations of motion.
Below, we simply list the basic choices.
\bqn
\{ f , I_1 , I_2 , I_3 \} = \left( \frac {-1}{\exp{(E)} } \right)
\frac {\partial (f , I_1 , I_2 , I_3 )}
{\partial (p_1 , p_2 , q_1 , q_2)}
\nqn

\bqn
\{ f , H , I_1 , I_2 \} = \left( \frac {-1}{2} \right)
\frac {\partial (f , H , I_1 , I_2 )}
{\partial (p_1 , p_2 , q_1 , q_2)}
\nqn

\bqn
\{ f , H , I_1 , I_3 \} = \left( \frac {-1}{2C_2 } \right)
\frac {\partial (f , H , I_1 , I_3 )}
{\partial (p_1 , p_2 , q_1 , q_2)}
\nqn

\bqn
\{ f , H , I_2 , I_3 \} = \left( \frac {1}{2C_1 } \right)
\frac {\partial (f , H , I_2 , I_3 )}
{\partial (p_1 , p_2 , q_1 , q_2)}
\nqn

Note that the constant factor in the above examples is usually
related to the integral of motion {\em not} chosen to be one
of the Hamiltonians (as was the case in the oscillator example
of Section 3). One can use the derivation property of the Nambu
bracket to find other brackets from the four listed above. For instance,
one can show that (37) and (38) are consistant with
\bqn
\{ f , H , I_1 , I_2 I_3 \} = \left (\frac {-1}{2(C_3 + C_2 ^{~2})} \right)
\frac {\partial (f , H , I_1 , I_2 I_3 )}
{\partial (p_1 , p_2 , q_1 , q_2)},
\nqn
and so on.

Finally, consider the (unphysical) Hamiltonian
\bqn
H = q_1 (p_1 - q_1 ) - q_2 ( p_2 - q_2 ),
\nqn
and its integrals of motion
\begin{eqnarray}
I_1 &=& q_1 (p_1 - q_1 ) = C_1 , \nonumber \\
I_2 &=& q_2 (p_2 - q_2 ) = C_2 , \\
I_3 &=& q_1 q_2 = C_3 . \nonumber
\end{eqnarray}
Since $ H=I_1 -I_2 $, there exists only the following basic Nambu bracket,
\bqn
\{ f , I_1 , I_2 ,I_3 \} = \left( \frac {1}{C_3 } \right)
\frac {\partial (f , I_1 , I_2 , I_3 )}
{\partial (p_1 , p_2 , q_1 , q_2)}.
\nqn
This two dimensional system can be extended to higher dimensions
by simply adding the appropriate terms to the Hamiltonian (41)
(such as $q_3 (p_3 - q_3 )$) and finding the extra integrals
of motion (i.e. $q_1 q_3 $ or $q_2 q_3 $).

\section {Conclusion}

We have demonstrated that several Hamiltonian systems possessing
dynamical symmetries can be realized in the Nambu formalism
of generalized mechanics. For all but one of these systems, an
extra freedom was found in the choice of the generalized
Hamiltonians needed for their Nambu construction. Finally, one
may speculate that since the harmonic oscillator is a very
important example in quantum mechanics, its Nambu formulation
may lead to a better understanding of the yet unsolved problem
of the quantization of Nambu mechanics.

\vspace{1cm}
{\bf Acknowledgements} \\
The author is grateful to L.A.Takhtajan for suggesting this
problem and for many helpful discussions. Thanks also go to
M.Flato for suggesting improvements in Section 4 of this
paper.\\

\section{References}

1. Nambu, Y., {\em Phys.~Rev.~D. } {\bf 7}, 2405 (1973).\\
2. Takhtajan, L.A., {\em Comm.~Math.~Phys. } {\bf 160}, 295 (1994). \\
3. Chakravarty, S., and Clarkson, P., {\em Phys.~Rev.~Lett. }
{\bf 65}, 1085 (1990). \\
4. Takhtajan, L.A., preprint PAM no.{\bf 121}, University
of Colorado (1991). \\
5. Arnold, V.I., {\em Mathematical Methods Of Classical Mechanics},
Springer-Verlag, Berlin, (1978). \\
6. Takhtajan, L.A., in Mo-Lin Ge and Bao-Heng Hao (eds), {\em
Introduction To Quantum Groups and Integrable Massive Models
of Quantum Field Theory }, World Scientific, Singapore, (1990). \\
7. Vaisman, I., {\em Lectures on the Geometry of Poisson
Manifolds}, Birkhauser, Berlin, (1994). \\
8. Sahoo, D., and Valsakumar, M.C., {\em Mod.~Phys.~Lett.~A.} {\bf 9},
2727, (1994). \\
9. Sahoo, D., and Valsakumar, M.C., {\em Phys.~Rev.~A. } {\bf 46},
4410 (1992). \\
10. Arnold, V.I., {\em Dynamical Systems 3 }, Springer-Verlag,
Berlin, (1988).

\end{document}